\documentstyle[prl,aps,psfig,preprint,tighten]{revtex}
\begin{document}

\draft

%\twocolumn [ \hsize\textwidth\columnwidth\hsize\csname
%@twocolumnfalse\endcsname

\title{Induced long range dipole field enhanced antihydrogen formation in the
$\bar{p}+ Ps(n=2)\rightarrow e^- + \bar{H}(n\le 2)$ reaction}
\author{Chi Yu Hu, David Caballero and Zolt\'an Papp }

\address{Department of Physics and Astronomy, 
California State University,
 Long Beach, California 90840}
\date{\today}
\maketitle

\begin{abstract}
\noindent
We assume all interaction to be Coulombic and solve the modified Faddeev
equation for energies between the $Ps(n=2)$ and $\bar{H}(n=3)$, which
involve six and eight open channels. We find that $99\%$ of the antihydrogen
are formed in $\bar{H}(n=2)$. Just above the $Ps(n=2)$ threshold
the S, P, and D partial waves contribute more than $4000$ square Bohr
radii near the maximum. Evidences indicate that the induced long range
dipole potential from the degenerate $Ps(n=2)$ targets is responsible
for such a large antihydrogen formation cross section.
\end{abstract}

\vspace{0.5cm} 

%]
\narrowtext
Due to the degeneracy of the excited hydrogen targets a long range dipole
potential is induced in the field of the incoming charged particle.
This potential is well known to be responsible for the formation of
Feshbach resonances just below the $n\ge 2$ thresholds \cite{1,2}.
In the absence of relativistic effects and low lying open channels
this attractive dipole potential can support an infinite set of bound states
just below the threshold. Relativistic corrections remove the degeneracy and
cut down the number of such states
to only a few. The presence of low lying open channels embedding them in the
continuum, they become resonant states and cause closely spaced oscillations
in the cross section just below the threshold.

Gilitis and Damburg \cite{2} pointed out that due to the Levinson theorem
\cite{3}
similar oscillations exist just above the threshold. Our S-state 
cross sections in the energy gap between the $Ps(n=2)$ and $\bar{H}(n=3)$
thresholds indeed reveal such oscillations and the antihydrogen formation 
cross section is greatly enhanced in this region slightly above the $Ps(n=2)$
threshold. It is interesting to point out that the locations of the
two Feshbach resonances below the $\bar{H}(n=3)$ threshold coincide almost
exactly to the two minimums of the antihydrogen formation cross section curve.
In addition, antihydrogen formation cross sections for  P and D partial waves
are calculated for a number of energies. It is shown that the total antihydrogen
formation cross sections from S, P and D partial waves
rise almost up to $1400 \pi \mbox{a}_0^2$ near the maximum region.

The calculations are carried out using the modified Faddeev equations. The
details of this method are given in Ref.\ \cite{4}. We provide an outline below.
The mass-scaled Jacobi vectors are defined as:
\begin{equation}
{\bf x}_\alpha=\tau_\alpha({\bf r}_\beta - {\bf r}_\gamma),\ \ \ 
{\bf y}_\alpha=\mu_\alpha \left({\bf r}_\alpha - 
\frac{m_\beta {\bf r}_\beta + m_\gamma {\bf r}_\gamma}{m_\beta+ m_\gamma }
\right), \label{1}
\end{equation}
where $\tau_\alpha=\sqrt{2 m_\beta m_\gamma/( m_\beta+m_\gamma)}$ and
 $\mu_\alpha=\sqrt{2 m_\alpha (1- m_\alpha / M) }$, $M= m_\alpha +
 m_\beta+m_\gamma$, and $(\alpha,\beta,\gamma)$ are cyclic permutations of
 $(1,2,3)$. The mass of the antiproton used is $m_1=1836.1527$, 
 the mass of the electron
 and the positron are equal, $m_2=m_3=1$ (the values are given in atomic units). 
 The Jacobi vectors of 
 different channels are related by the orthogonal transformation:
\begin{equation}
\left( 
\begin{array}{c}
{\bf x}_\beta \\
{\bf y}_\beta
\end{array} \right) =\left( 
\begin{array}{cc}
C_{\beta \alpha} & S_{\beta \alpha}  \\
-S_{\beta \alpha} & C_{\beta \alpha}
\end{array} \right)
\left( 
\begin{array}{c}
{\bf x}_\alpha \\
{\bf y}_\alpha
\end{array} \right),
\end{equation}
where 
\begin{eqnarray}
C_{\beta \alpha} & = & \left[ \frac{m_\beta m_\alpha}
{(M - m_\beta)(M- m_\alpha)}
\right]^{1/2},\\ S_{\beta \alpha} & = & (-1)^{\beta-\alpha} 
\mbox{sgn}(\alpha-\beta) (1-C_{\beta \alpha}^2)^{1/2}.
\end{eqnarray}

In our particular case  $V_3$, the interaction between the $\bar{p}$ and
$e^-$, is a repulsive Coulomb
potential which does not support two-body bound states. There are no
asymptotic channels associated with this fragmentation. Consequently
the total three-body wave function can be expressed in two components
\begin{equation}
\Psi=\psi_1 ( {\bf x}_1 ,{\bf y}_1) + \psi_2 ( {\bf x}_2 ,{\bf y}_2). 
\end{equation}
Asymptotically $\psi_1$ consists of an electron bound to the positron and a free
antiproton, while $\psi_2$, the rearrangement Faddeev component describes
asymptotically an positron bound to the antiproton and a free
electron.

We use the modified Faddeev equations:
\begin{eqnarray}
(-\Delta_{x_\alpha} -\Delta_{y_\alpha} + V_\alpha +  \bar{V}_\alpha -E) &&
\psi_\alpha ({\bf x}_\alpha ,{\bf y}_\alpha) =   \nonumber \\
&& - V_\alpha^{(s)}
\psi_\beta ({\bf x}_\beta ,{\bf y}_\beta),
\end{eqnarray}
$\bar{V}_\alpha=V_3+V_\beta^{(l)}$, $\alpha\ne\beta=1,2$, 
and $V_1$ and $V_2$, the interactions between the $(e^-,e^+)$ and $(\bar{p},e^+)$
pairs, respectively, are  
separated into short- and long-range terms
\begin{eqnarray}
V_\alpha^{(s)}&=& V_\alpha (x_\alpha) \zeta_\alpha (x_\alpha,y_\alpha), \\
V_\alpha^{(l)}&=& V_\alpha (x_\alpha) (1- \zeta_\alpha (x_\alpha,y_\alpha)).
\end{eqnarray}
The function $\zeta_\alpha(x_\alpha,y_\alpha)$ vanishes asymptotically 
within the three-body sector, where $x_\alpha \sim y_\alpha \to \infty$,
and approaches one in the two-body cluster region, where 
$ x_\alpha << y_\alpha \to \infty$. We use the following function
having the required property:
\begin{equation}
\zeta(x,y)=2 \left\{ 1+ \exp \left[ (x/x_0)^\nu /(1+y/y_0) \right]
\right\}^{-1},
\end{equation}
where $\nu$ must be larger than 2 \cite{4}. In principle, $x_0$ and $y_0$
are arbitrary, but they should be chosen to be consistent with
the size of the scattering system for a rapid convergence.

In bipolar basis \cite{5} the  Faddeev component
is given in the form
\begin{equation}
\Psi_\alpha ( {\bf x}_\alpha ,{\bf y}_\alpha)=
    \sum_{L=0}^{\infty}  \sum_{M=-L}^L 
\sum_{\vec{l}+\vec{\lambda}=\vec{L}} 
\frac{\psi_{\alpha_{l \lambda}}^L 
({x}_\alpha, {y}_\alpha) }{{x}_\alpha {y}_\alpha} 
Y_{l \lambda}^{L M} (\hat{x}_\alpha, \hat{y}_\alpha) 
\end{equation}
where
$Y_{l \lambda}^{L M} (\hat{x}_\alpha, \hat{y}_\alpha)=\left[
Y_{l}^{m_l} (\hat{x}_\alpha ) \times Y_{\lambda}^{m_\lambda} (\hat{y}_\alpha )
\right]$, $\alpha=1,2$,  and  $\psi_{\alpha_{l \lambda}}^L 
({x}_\alpha, {y}_\alpha)$ is the partial component of the three-body wave
function having total angular momentum $L$ and relative angular momenta
$l$ and $\lambda$ associated with coordinates ${\bf x}_\alpha$ and
${\bf y}_\alpha$, respectively. Thus, for each $L$, the MFE \cite{5} is further
reduced to a set of two-dimensional partial differential equations:
\widetext
\begin{eqnarray}
\lefteqn {(H_l^{(\alpha)}+V_\alpha -E)
\psi^L_{\alpha_{l\lambda}} ({x}_\alpha, {y}_\alpha) 
+ \sum_{\vec{l}'+\vec{\lambda}'=\vec{L} } 
W_{l\lambda, l'\lambda'}^{(\alpha)L}( {x}_\alpha, {y}_\alpha)
\psi_{\alpha_{l'\lambda'}}^L  ({x}_\alpha, {y}_\alpha) = }
\nonumber \\
&& -V^{(s)}_\alpha (x_\alpha,y_\alpha) 
\sum_{\vec{l}'+\vec{\lambda}'=\vec{L}} 
\langle Y_{l \lambda}^{L M} (\hat{x}_\alpha, \hat{y}_\alpha)|
\frac{\psi_{\beta_{l' \lambda'}}^L 
({x}_\beta, {y}_\beta) }{{x}_\beta {y}_\beta} 
Y_{l' \lambda'}^{L M} (\hat{x}_\beta, \hat{y}_\beta) \rangle,
\end{eqnarray}
where
\begin{equation}
H_l^{(\alpha)}=-\partial^2_{x_\alpha}-\partial^2_{y_\alpha} 
+ \frac{l(l+1)}{x^2_\alpha}+\frac{\lambda (\lambda+1)}{y^2_\alpha},
\end{equation}
\begin{equation}
W_{l \lambda, l' \lambda'}^{(\alpha)L}  ( {x}_\alpha, {y}_\alpha)  =
\langle Y_{l \lambda}^{L M} (\hat{x}_\alpha, \hat{y}_\alpha)|
\bar{V}_\alpha
| Y_{l' \lambda'}^{L M} (\hat{x}_\alpha, \hat{y}_\alpha) \rangle,
\end{equation}
and  $\beta\ne\alpha=1,2$.
 
\narrowtext
The summation with respect to angular momentum channels is 
truncated in Eq.\ (11). When the proper cut-off parameters $x_0$ 
and $y_0$ in (9) is chosen the number of terms used for the S, P, and D
partial waves ranges from $10$ to $12$ for the first Faddeev channel 
($\bar{p}+Ps$), $8$ to $10$ for the second Faddeev channel ($e^- +\bar{H}$).
The optimal choices of parameters satisfy the conditions 
$x_{\alpha_{max}} \approx 10 x_0$; $y_{\alpha_{max}} \approx 10 y_0$, $\alpha=1,2$,
where $x_{\alpha_{{max}}}$, $y_{\alpha_{{max}}}$ are the respective 
cut-off distances which divide the asymptotic and interior regions. 
In the energy gap between $Ps(n=2)$ and $\bar{H}(n=3)$ we used 
$110\mbox{a}_0$ and $75\mbox{a}_0$ for $x_{\alpha_{{max}}}$, $\alpha=1,2$, and
 $450\mbox{a}_0$ and $150\mbox{a}_0$ for $y_{\alpha_{{max}}}$, respectively.
 
In the MFE approach 
the total three-body wave function $\psi^{(\sigma)}(x,y)$, for each
open channel $\sigma$, is a vector having all $\psi^L_{\alpha_{l\lambda}}$
as its components. We further split the wave function into interior and
asymptotic parts using the vector equation
\begin{equation}
\psi^{(\sigma)}(x,y)=F^{(\sigma)}(x,y)+ f^{(as)}_{\sigma}(x,y) +
\sum_{\sigma'=1}^{\sigma_{max}} 
\tilde{K}_{\sigma' \sigma} f_{as}^{(\sigma')}(x,y),
\end{equation}
where $F^{(\sigma)}(x,y)|_{y_\alpha > y_{\alpha_{max}}} \equiv 0$,
$f^{(as)}_{\sigma}(x,y)$ and $f_{as}^{(\sigma')}(x,y)$ are known incoming and
outgoing asymptotic wave functions for the open channels $\sigma$
when $y_\alpha > y_{\alpha_{max}}$ and they are equipped with spline continuity
of values, first and second derivatives across the boundaries.
They are otherwise identically zero in the interior regions. 
$\tilde{K}_{\sigma' \sigma}$ is related to the K-matrix by a kinematic factor
\cite{6}.

The partial cross sections between the incoming channel $i$ and the outgoing
channel $j$ with total angular momentum $L$ is given by
\begin{equation}
\sigma_{i j}=\frac{\pi \mbox{a}_0^2}{k_i^2} (2 L +1 )
\left| \left(\frac{2K}{1-\mbox{i}K}\right)_{i j} \right|^2 ,
\end{equation}
where $k_i$ is the momentum of the incoming channel.

Upon substitution of (14) into (11), we obtain the vector equation
\begin{equation}
(H-E) F^{(\sigma)}(x,y) = I_\sigma + \sum_{\sigma'=1}^{\sigma_{max}}
\tilde{K}_{\sigma' \sigma} I^{(\sigma')}.
\end{equation}
$H$ is composed of two diagonal blocks of operators corresponding to the two 
Faddeev components on the left-hand side of (11) and two
off-diagonal rectangular blocks involving the short-range potentials on the
right-hand side of (11). $I_\sigma$, $I^{(\sigma')}$ are known inhomogeneous
column vectors. Replace the unknown vector $F^{(\sigma)}(x,y)$
by  $F^{(\sigma)}(x,y)=U_\sigma (x,y)+\sum_{\sigma'=1}^{\sigma_{max}}
\tilde{K}_{\sigma',\sigma} U^{(\sigma')}$ in (16). We find the $U$'s satisfy
the inhomogeneous equations
\begin{eqnarray}
(H-E) U_\sigma & = & I_\sigma \\
(H-E) U^{(\sigma)} & = & I^{(\sigma)}, \ \ \ \ \sigma=1,\ldots,\sigma_{max}. 
\end{eqnarray}
The $U$'s are solved using fifth order Hermite polynomial spline expansion
\cite{5,6,7}. Matching the interior wave function $F^{(\sigma)}(x,y)$ with the
known asymptotic wave functions at or near $y_{\alpha_{max}}$ we obtain
highly over determined linear equations for the unknown 
$\tilde{K}_{\sigma' \sigma}$. Two independent methods are used for the
solutions: (1) a standard least-square procedure and (2) projection
on open channels. The details are given in \cite{5,6}.

The size of matrix equation used to solve (17,18) varies depending on the energy
and the total angular momentum.  They range from about $100000\times 100000$
to  $150000\times 150000$ for results accurate at least $5\%$. 
The solutions are
obtained using a solver in ScaLAPACK library on the massive parallel 
Blue Horizon IBM computer at the San Diego Supercomputer Center.

\begin{figure}
\psfig{file=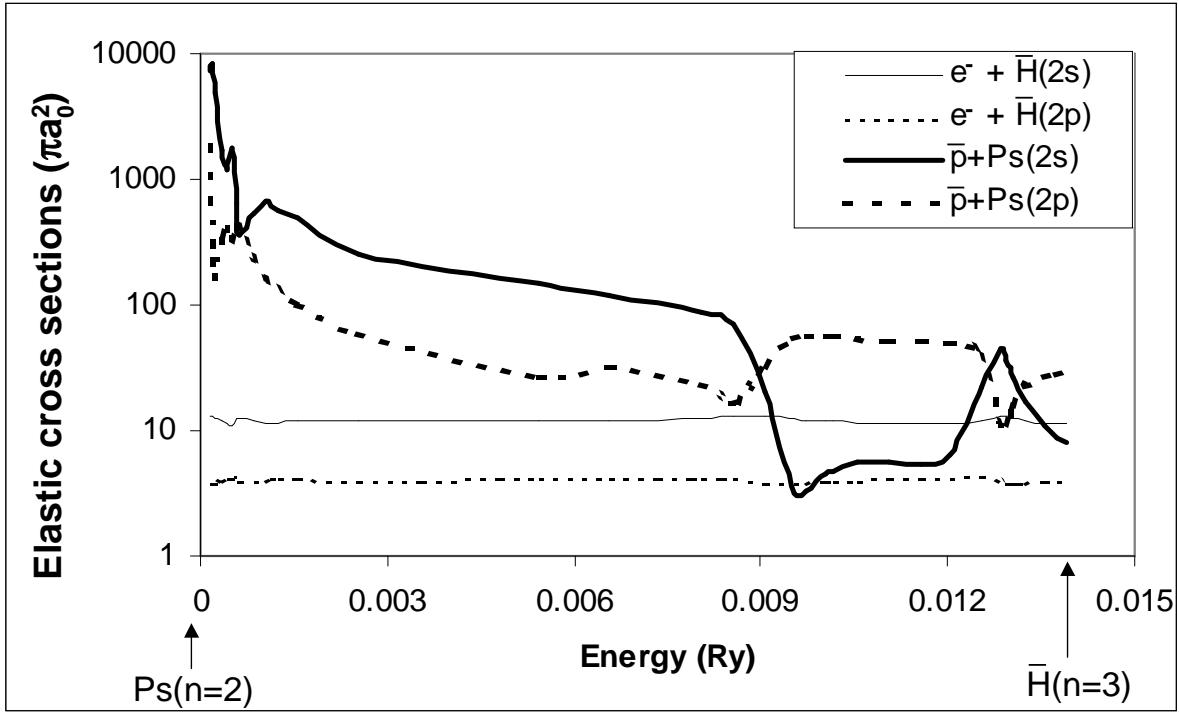,width=16cm}
\caption{Elastic cross sections of the processes 
$e^- + \bar{H}(2s)$, $e^- + \bar{H}(2p)$,
$\bar{p}+Ps(2s)$  and $\bar{p}+Ps(2p)$.
All energies are measured from the $Ps(n=2)$ threshold. The $\bar{H}(n=3)$
threshold is located at $0.01395 Ry$.}
\label{fig1}
\end{figure}

In the energy gap between $Ps(n=2)$ and $\bar{H}(n=3)$ there are six open
channels for S partial wave. The elastic cross sections are plotted
in Fig.\ \ref{fig1}, the horizontal axis 
plots energies in $Ps(n=2)+\bar{p}$ channel.
The elastic cross section from $e^- + \bar{H}(1s)$ and $\bar{p}+Ps(1s)$
are relatively very small, they are not visible, that from 
$e^- + \bar{H}(2s)$ $e^- + \bar{H}(2p)$ are somewhat larger. They are plotted 
in Fig.\ \ref{fig1}, for comparison with the huge elastic cross sections
from the channels $\bar{p}+Ps(2s)$ and $\bar{p}+Ps(2p)$. Due to the large
induced dipole moment of the excited positronium, the Gilitis-Damburg 
oscillations are clearly seen in Fig.\ \ref{fig1} below $0.003 \mbox{Ry}$,
the energies are measured from the $Ps(n=2)$ threshold. 
 The cross sections are rather monotonous between 
$0.003-0.008 \mbox{Ry}$. The structure below the $\bar{H}(n=3)$ threshold is
due to the two $S$-state resonances located at $0.8835 \mbox{Ry}$ 
$0.8875 \mbox{Ry}$ measured from the $\bar{H}(1s)$ threshold,
or $0.0090 \mbox{Ry}$ and $0.0130 \mbox{Ry}$ relative to $Ps(n=2)$ threshold.
One notices that the oscillations of the two degenerate $2s$ and $2p$ channels
are nearly $180^\circ$ out of phase in the region of Feshbach resonances, although
the effect is distorted by the statistical weight of $1/3$ on the $2p$ cross
sections.

\begin{figure}
\psfig{file=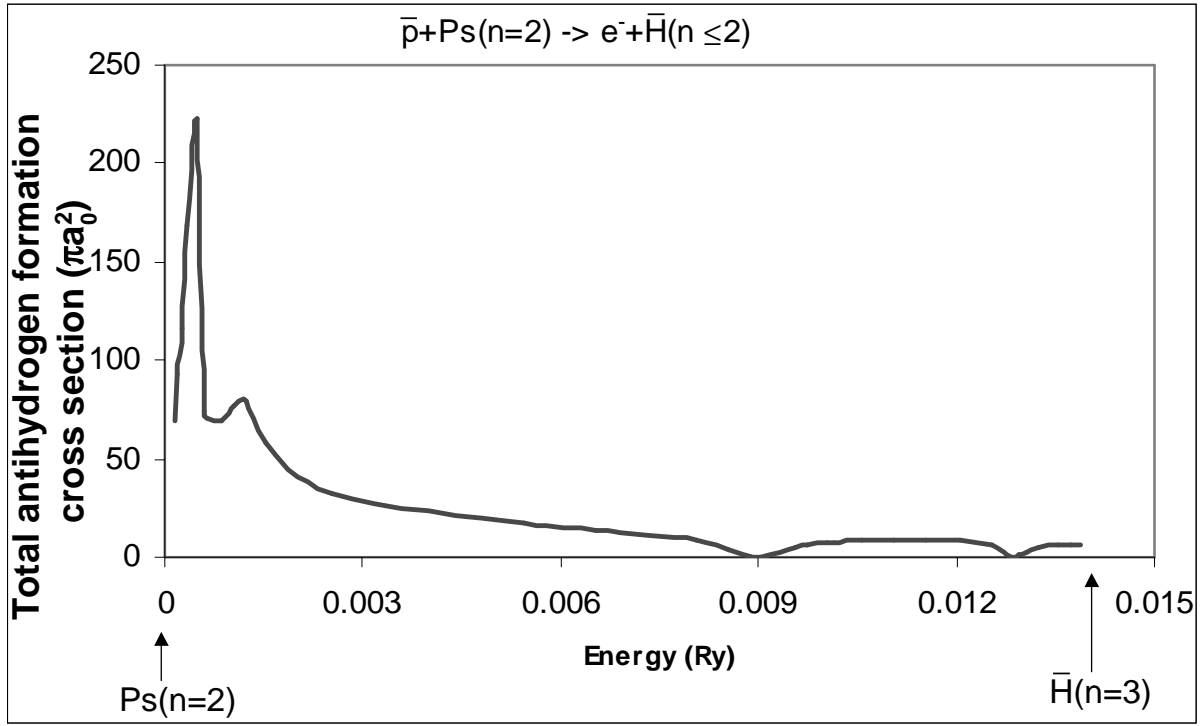,width=16cm}
\caption{Antihydrogen formation cross sections from all the processes
$\bar{p}+Ps(n\le 2) \rightarrow e^- + \bar{H}(n\le 2)$. The energies are
measured from the $Ps(n=2)$ threshold.}
\label{fig2}
\end{figure}

Fig.\ \ref{fig2} presents the $S$-state antihydrogen 
formation cross sections in the same
energy gap. The formation cross section are greatly enhanced in the oscillation
region below $0.003 \mbox{Ry}$. The largest cross section calculated is 
$219 \pi \mbox{a}_0^2$. The two minima below the $\bar{H}(n=3)$ threshold are
located almost exactly at the location of the two 
Feshbach resonances \cite{8,9}.

Calculations of the $P$ and $D$ partial wave cross sections have been carried
out for a number of energies in the oscillating region. There are 
$8$ open channels
and $64$ partial cross sections, most of them will not be reported here.
Table I presents all antihydrogen formation cross sections of $S$, $P$ and 
$D$ partial waves at four energies. There are $9$ antihydrogen formation
partial cross sections for $S$-wave and $16$ such partial cross sections
for $P$ and $D$ waves, respectively. We group them into three partial sums,
$\sigma_L^{(1)}$, $\sigma_L^{(2)}$ and $\sigma_L^{(3)}$, where $L$ is the
angular momentum $(S,P,D)$. $\sigma_L^{(1)}$ is the sum of all partial 
cross sections with antihydrogen formed in $1s$ state. 
$\sigma_L^{(2)}$ is the sum of all partial cross sections having antihydrogen
formed in $2s$ and $2p$ states but originated from $Ps(1s)$ target.
$\sigma_L^{(3)}$ is the sum of all partial cross sections with antihydrogen
formed in $2s$ and $2p$ states but originated from excited $Ps$ targets.

Our calculations highlight three important points:
\begin{enumerate}
\item The $P$ and $D$ partial waves follow the same tendency of strong
oscillations near $Ps(n=2)$ threshold similarly to that of $S$-partial wave.
Table I shows as $\sigma_S$ increases towards the $Ps(n=2)$ threshold, so do 
$\sigma_P$ and $\sigma_D$. Seaton \cite{1} pointed out that the strength of the 
dipole potential from $n=2$ target can support Feshbach resonances for all
partial waves with $L\le 2$.

\item Since the cross section depend on the kinematic factor $1/k_i^2$
in (15), where $k_i$ is the momentum of the incoming channel, the oscillations
are amplified near the lower threshold $Ps(n=2)$, and hence the
antihydrogen formation cross sections are also greatly enhanced. On the other
hand, the
Feshbach resonances just below the upper $\bar{H}(n=3)$ threshold
 have negligible contribution to the antihydrogen formation cross sections.
In fact, Fig.\ \ref{fig2} shows two minima in antihydrogen formation cross section
located at
$0.8835 \mbox{Ry}$ and $0.8875 \mbox{Ry}$ above $\bar{H}(1s)$ threshold, 
respectively. Archer and Parker \cite{8} found one $S$-state resonance at 
$-0.11492 \mbox{Ry}$ below breakup, or $0.88454 \mbox{Ry}$ above $\bar{H}(1s)$.
Ho and Green \cite{9} found two $S$-state resonances at $-0.11606  \mbox{Ry}$
and $-0.11206  \mbox{Ry}$ below breakup, 
or $0.88340 \mbox{Ry}$ and $0.88740 \mbox{Ry}$ above $\bar{H}(1s)$.
We used the threshold energy $-0.99946 \mbox{Ry}$ for $\bar{H}(1s)$,
due to finite $\bar{p}$ mass, to calculate all resonance positions.
In comparison, our novel method to determine Feshbach resonance
position seems to have at least three to four digits of accuracy!

\item Not surprisingly, the long range polarization potential, mainly the 
dipole potential, couples strongly only amongst the excited $2s$ and $2p$ 
states which lie just below the targets $Ps(n=2)$. 
The ground states $Ps(1s)$ and $\bar{H}(1s)$ contribute
 less than $2\%$ of the antihydrogen formation cross section,
and even less of them are formed in $\bar{H}(1s)$.
As a result, near $99\%$ of the antihydrogen are formed in $2s$ and $2p$
states with $Ps(n=2)$ targets.
\end{enumerate}

Archer and Parker \cite{8} solved the Schr\"odinger equation using a form of
hyperspherical coordinates. They calculated $S$-state partial cross sections up
to $H(n=4)$ threshold. However, they purposely avoided the long range 
polarization potentials by projecting out only those with ground states
targets and using cut-off hyperradius of $\rho=120.0 \mbox{a}_0$.
We used cut-off coordinates that are equivalent to
 $\rho\approx 470 \mbox{a}_0$.  Our corresponding
partial cross sections in the energy-gap between $Ps(n=2)$ and $\bar{H}(n=3)$
are generally $25\%$ lower.  The complete coupling of all six open channels
in our calculation may be responsible for the differences.

Igarashi et.\ al.\ \cite{10} made hyperspherical coupled-channel calculations
for antihydrogen formation in a broad range of energies in $\bar{p}+Ps$
scattering including partial waves up to angular momentum $L=12$ and included
cross section from Born approximations for $L\ge 13$. Unfortunately
there is only one of their energy points fell within the gap between
$Ps(n=2)$ and $\bar{H}(n=3)$ thresholds. This point is located at 
$\sim 0.002 \mbox{Ry}$, its total formation cross section is approximately
$640 \pi \mbox{a}_0^2$, where our value at this point is approximately
$396 \pi \mbox{a}_0^2$ from three partial waves. This means that, at least at
this energy, the lowest three partial waves contribute about $62\%$
of the total formation cross section.  Fig.\ \ref{fig2} and Table I show that this
energy point is not in the region of optimal antihydrogen formation.
Our largest calculated antihydrogen formation cross section is 
$1397 \pi \mbox{a}_0^2$ from the lowest three partial waves. The energy at this 
point is $0.000494 \mbox{Ry}$. Table I shows that at this energy
$\sigma_D/\sigma_S = 3.15$. This is less then the factor $2L+1$ in (15). But it
is expected that higher partial waves contribute much less effectively at 
such low energy. The long range dipole potential from $Ps(n=2)$ targets is
expected to cause oscillations for all partial waves with $L\le 2$.
We expect a much more drastic drop in $\sigma_L/\sigma_S$ for $L\ge 3$.
Works are in progress to calculate contributions from $L\ge 3$.

\acknowledgements
This work has been supported by the NSF Grant No.Phy-0088936.
We also acknowledge the
generous allocation of computer time at the NPACI, formerly 
San Diego Supercomputing Center.

\begin{table}[tbp]
\caption{Antihydrogen formation cross sections and partial cross sections for 
processes $\bar{p}+Ps(n\le2)\rightarrow e^- + \bar{H} (1s)$,
$\bar{p}+Ps(n=1)\rightarrow e^- + \bar{H} (n=2)$ and
$\bar{p}+Ps(n=2)\rightarrow e^- + \bar{H} (n=2)$. The corresponding partial 
cross sections are denoted by  $\sigma_L^{(1)}$,
$\sigma_L^{(2)}$ and $\sigma_L^{(3)}$, respectively, and $L=S,P,D$.
The energy values (in Ry)  are measured from the $\bar{H}(1s)$ threshold 
and the cross sections are given 
in $ \pi a_0^2$ units. $\sigma_L = \sum_{i}\sigma_{L}^{(i)}$ is the total
formation cross section for the partial wave $L$. }
\label{4channel}
\begin{tabular}{|c|cccc|}  
   & \multicolumn{4}{c|}{ Energy } \\ 
    &   0.8749  &   0.8755   &   0.877   &   0.8799   \\ \hline
$\sigma_{S}^{(1)}$ &   0.282   &   0.097    &   0.047   &   0.030    \\  
$\sigma_{S}^{(2)}$ &   0.125   &   0.116    &   0.112   &   0.107    \\  
$\sigma_{S}^{(3)}$ &  218.84   &   76.701   &   32.481  &  17.201    \\  \hline
$\sigma_{S}$
               &  219.25   &  76.914    &   32.640  &  17.338  \\ \hline
$\sigma_{P}^{(1)}$ &   3.373   &   1.783    &   1.130   &   0.886    \\  
$\sigma_{P}^{(2)}$ &   1.041   &   1.042    &   1.015   &   1.040    \\  
$\sigma_{P}^{(3)}$ &  482.65  &  226.62   &  101.91  &  50.73   \\  \hline
$\sigma_{P}$
               &  487.06  &  229.45   &  104.06  &  52.66  \\ \hline
$\sigma_{D}^{(1)}$ &   7.841   &   4.214    &   2.668   &   2.087    \\  
$\sigma_{D}^{(2)}$ &   1.076   &   1.068    &   1.047   &   1.099    \\  
$\sigma_{D}^{(3)}$ &  682.02  &  312.94   &  162.91  &  76.174   \\  \hline
$\sigma_{D}$
               &  690.94  &  318.22   &  166.63  &  79.36  \\ \hline
$\sum_{L=S,P,D}\sigma_{L}^{(1)} $ &  11.496   &   6.094    &   3.845   
&   3.003    \\  
$\sum_{L=S,P,D}\sigma_{L}^{(2)}  $ &   2.242   &   2.226    &   2.174   
&   2.246    \\  
$ \sum_{L=S,P,D}\sigma_{L}^{(3)} $ & 1383.51  &  616.26   &  297.31  
& 144.11   \\  
\hline
$\sum_{L=S,P,D} \sigma_{L}   $
               &  1397.25 &  624.58   &  303.33  &  149.36 
\end{tabular}
\end{table}

\newpage

\newpage

\end{document}